\newcommand{\cm}{\rm \,cm}
\newcommand{\mm}{\rm \,mm}
\newcommand{\m}{\rm \,m}

\newcommand{\km}{\rm \,km}
\newcommand{\e}{\rm e}
\newcommand{\s}{\rm \,s}

\newcommand{\sr}{\rm \,sr}

\newcommand{\MeV}{\rm \,MeV}
\newcommand{\eV}{\rm \,eV}

\newcommand{\GeV}{\rm \,GeV}
\newcommand{\TeV}{\rm \,TeV}

\newcommand{\anue}{$\bar{\nu}_e$}

\newcommand{\lsim}{\lower .5ex\hbox{$\buildrel < \over {\sim}$}}
\newcommand{\gsim}{\lower .5ex\hbox{$\buildrel > \over {\sim}$}}

\newcommand{\stitolo}{
Neutrino Physics and Astrophysics with the MACRO Experiment at the 
Gran Sasso Lab 

}
\newcommand{\scollabor}{

}
\newcommand{\sautori}{

Giorgio~Giacomelli\footnote{MACRO Collaboration. See Ref. 1 for a list of MACRO Collaborators and Institutions.}
 
}
\newcommand{\sfirmatoda}{

\collabor                            

}
\newcommand{\sistituzioni}{
                                      
Dipartimento di Fisica dell'Universit\`{a} di Bologna and INFN, Sezione di Bologna, 
I-40127 Bologna, Italy, e-mail: giacomelli@bo.infn.it \\
\vspace{0.3cm}
\bf Invited paper at the \(25^{th}\) Meeting of the Nuclear Division of the Brasilian Physical Society, S. Pedro, Brasil, 1-4 September 2002}

\newcommand{\sabst}{The results of the MACRO experiment on atmospheric neutrino 
oscillations are presented and discussed. The data concern different event topologies 
with average neutrino energies of \(\sim\) 4 and \(\sim\) 50 GeV. The Multiple Coulomb Scattering of muons in the MACRO absorbers was used to 
estimate the neutrino energy of each event of the higher energy sample. The angular 
distributions, the absolute fluxes and the L/E distribution strongly favour 
\(\nu_{\mu} \rightarrow \nu_{\tau}\) oscillations with maximal mixing and 
\(\Delta m^{2}=0.0025 \; eV^{2}\).
Results are presented on the searches for astrophysical sources of high energy muon 
neutrinos, for bursts of electron antineutrinos from stellar gravitational collapses 
and on indirect searches for WIMPs from the Earth and from the Sun.}

\documentclass[12pt,twoside]{article}
\usepackage{epsfig}
\newcommand{\preprintnum}{
\begin{flushright}
{\bf DFUB 9/2002 }\\
{\bf 21/9/2002}\\ 
\end{flushright}}
\newcommand{\titolo}{
\begin{center}\Large{\bf{
\stitolo
}}\end{center}}
\newcommand{\collabor}{
\begin{center}
\Large{
\scollabor                             
}    
\end{center}}
\newcommand{\autori}{
\begin{center}
\sautori
\end{center}}
\newcommand{\istituzioni}{
\begin{center}
\small{\it{
\sistituzioni
}}                       
\end{center}}                         
\newcommand{\abst}{
\begin{abstract}
\sabst               
\end{abstract}}
\newcommand{\firmatoda}{
\begin{center}
\sfirmatoda
\end{center}}
\begin{document}

%
\preprintnum
\titolo
\firmatoda
\autori
\istituzioni
\abst
%
\setcounter{page}{1} 
\pagestyle{plain}                
%

%
\section{Introduction}

MACRO was a large area multipurpose underground experiment designed
to search for rare events in the penetrating cosmic radiation. 
These included the study of atmospheric
neutrinos and their oscillations, high energy 
\( (E_{\nu }\, \, \gsim \, \, 1\, \, \GeV ) \) muon 
neutrino astronomy, indirect searches for WIMPs and search for low energy
(\( E_{\nu }\, \, \gsim \, \, 7\, \,  \)MeV) stellar collapse neutrinos.
The detector was placed in the Gran Sasso laboratory, located on the highway 
Rome-Teramo , 120 km east of Rome. The lab consists of three underground halls, each about 100 m long. It is at an altitude of 963 m above sea level, is well 
shielded from cosmic rays by a mean rock thickness of \( \simeq 3700 \) m.w.e.; the minimum is \( 3150 \) m.w.e: this defines the minimum muon
energy at the surface at \( \sim 1.3{\TeV } \) in order to reach
MACRO. The average residual muon energy and the muon flux are 
\( \sim 320{\GeV } \) and \( \sim 1{\m }^{-2}{\textrm{h}}^{-1} \),
respectively. 

MACRO was composed of three sub-detectors: liquid scintillation
counters, limited streamer tubes and nuclear track detectors. Each
one of them could be used in \lq\lq stand-alone\rq\rq~and in \lq\lq combined\rq\rq~mode.
A cross section of the detector is shown in Fig. 1. Notice
the division in the \textit{lower} and in the  \textit{upper} part (this was 
often
referred to as the \textit{Attico}); the inner part of the \textit{Attico} was
empty and lodged the electronics [1]. The mass of the \textit{lower}
MACRO was \( \simeq 4200\, \,  \)t, mainly in the form of boxes filled with
crushed Gran Sasso rock.
The detector had a modular structure: it was divided into six
sections referred to as supermodules. Each active part of one supermodule
had a size of \( 12.6\times12 \times9 .3{\m }^{3} \) and had a separate
mechanical structure and electronics readout. The full detector had
global dimensions of \( 76.5\times12 \times9 .3{\m }^{3} \) and provided
a total acceptance to an isotropic flux of particles of 
\( \sim 10000{\m }^{2}{\sr }. \) The total mass was \( \simeq 5300\, \,  \)t.



\begin{figure}
\vspace{-1cm}
  \begin{center}
  \mbox{ 
\epsfysize=8cm
         \epsffile{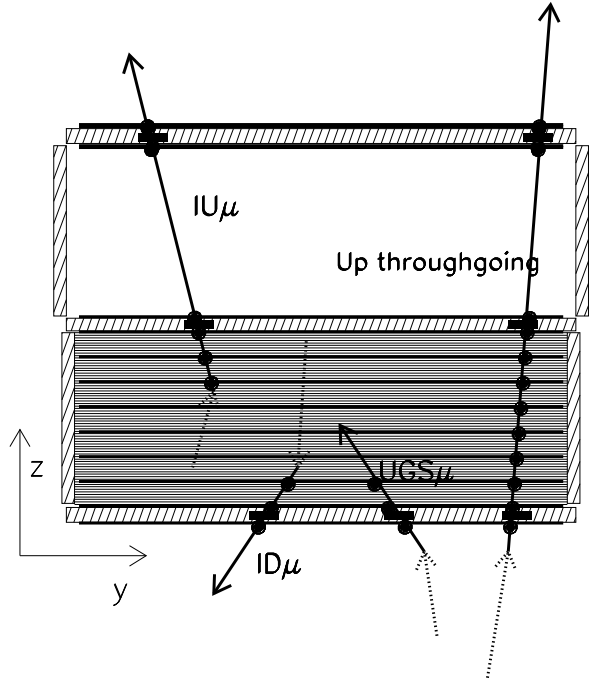} 
}   
\caption{\label{fig2}\small Cross section of the MACRO detector and sketch of 
different event
topologies induced by $\nu_\mu \rightarrow \mu $ charged current interactions.
The black points and the black rectangles represent streamer tubes and
scintillator hits, respectively. Tracking was performed by the streamer
tubes;  the time-of-flight (T.O.F.) of the muons was measured by the
scintillators for {\it Up Semicontained } (Internal upgoing - IU $\mu$)
and {\it Upthroughgoing }  events (and also for downgoing muons).}
\end{center}
\end{figure}
Data were taken from March 1989 till april 1994 with the detector under construction and from the middle of 1994 till the end of 2000 with the full detector.

\section{Atmospheric  neutrino oscillations}
If neutrinos have non-zero masses, one has to consider the 
{ \it weak flavour eigenstates} \(\nu_{\e}, \nu_{\mu}, \nu_{\tau}\) and the 
{\it mass eigenstates} \(\nu_{1}, \nu_{2}, \nu_{3}\). The weak flavour 
eigenstates  \(\nu_{\l}\) are linear combinations of the mass eigenstates 
\(\nu_{\m}\) through the elements of the mixing matrix  \( U_{lm} \):
\begin{equation}
\nu_{\l} = \sum_{m=1}^{3}\; U_{lm} \;  \nu_{m}
\end{equation}
In the simple case of only two flavour eigenstate neutrinos 
(\(\nu_{\mu}, \nu_{\tau}\)) which oscillate with two mass eigenstates 
(\(\nu_{2}, \nu_{3}\)) one has

\begin{equation}
\left
\{ \begin{array}{lc} 
\nu_{\mu} = &\nu_{2} \cos \theta_{23} + \nu_{3} \sin  \theta_{23} \\
\nu_{\tau}  = &-\nu_{2}\sin \theta_{23} + \nu_{3}\cos \theta_{23} 
\end{array} 
\right. 
\end{equation}

where  \(\theta_{23}\) is the mixing angle (\(\theta\) will be used in the following). In this case one may easily compute the 
following expression for the survival probability of a \(\nu_{\mu}\) beam

\begin{equation}
P(\nu_{\mu}\rightarrow \nu_{\mu})=1 - \sin^{2} 2\theta_{23} \sin^{2} 
\left( \frac{E_{2} - E_{1}}{2} t \right) = 1 - \sin^{2} 2\theta_{23} \sin^{2}
\left( \frac{1.27 \Delta m^{2} \cdot L}{E_{\nu}}
\right)
\end{equation}
where \(\Delta m^{2}=\m^{2}_{3} - \m^{2}_{2}\) and L is the distance travelled by the muon neutrino from 
production to detection.

High energy primary cosmic rays interact in the upper atmosphere producing pions and Kaons, which by decay yield muons and muon neutrinos, 
\(\pi \rightarrow \mu \nu_{\mu}\); further decays of the muons lead to electron and muon neutrinos, \(\mu \rightarrow \nu_{\mu} \nu_{\l} \l\). The neutrinos 
are produced in a spherical shell at about 10- 20 km from the earth surface. 
Upgoing \(\nu_{\mu}\)'s may interact in the rock below MACRO or inside its lower part leading to upgoing muons, \(\nu_{\mu} N \rightarrow \mu^{+-} +...\).

 Upward going muons are identified by the streamer tube system (for
tracking) and the scintillator system (for time-of-flight measurement).
A rejection factor of at least \( 10^{7} \) is needed in order to
separate upgoing muons from the  background due to the
downgoing muons. Fig. \ref{fig2} shows sketches of the  different  neutrino 
event
topologies  analyzed: Upthroughgoing muons,
Upsemicontained (also called Internal Upgoing muons, IU),  Upgoing Stopping
muons (UGS), Internal Downgoing muons (ID). The numbers of events 
measured and expected for the three topologies are
given in Table \ref{tab:macro}. All data samples deviate from MC
expectations; the deviations point to neutrino oscillations.

\subsection{Upthroughgoing muons}

The \textit{upthroughgoing muons} come from \( \nu _{\mu } \) interactions
in the rock below the detector; the \( \nu _{\mu } \)'s have an average
energy \( \overline{E}_{\nu }\sim \, \, 50\, \, \GeV  \) [2, 3]. The data 
 refer to a livetime of  6.16 years (full detector 
equivalent). The data deviate in absolute value and in shape from Monte Carlo 
(MC) predictions, see Fig. 2. 
We studied a large number of possible effects that could
affect our measurements:  no significant systematic
problems exist in the detector or in the data analyses. 

\begin{figure}[t]
 \vspace{-1.cm}
 \begin{center}
  \mbox{ \epsfysize=8cm
         \epsffile{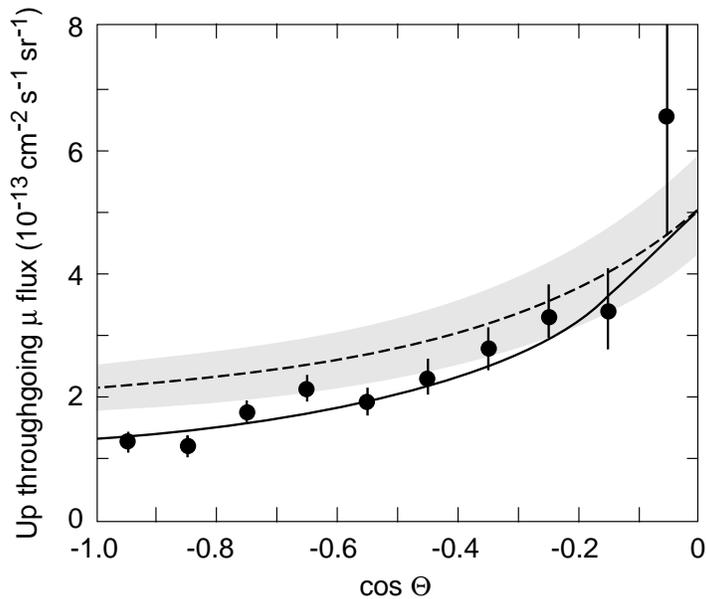}   }
 \end{center}
\vspace{-.3cm}
\caption{\label{fig6}\small Zenith angle distribution of upthroughgoing muons (black points).
The dashed line is the expectation for no oscillations (with a 17 \% scale uncertainty band). The solid line is the fit
for an oscillated muon flux which yields maximum mixing and $\Delta m^{2} = 
2.5\cdot 10^{-3}$ eV$^{2}$ [2] [3].}
\end{figure}

The measured data have been compared with MC simulations.
For the upthroughgoing muon simulation, the neutrino flux computed
by the Bartol group was used [4]. The cross sections for the neutrino interactions
were calculated using the deep inelastic parton distributions of ref. [5].
The muon propagation to the detector was done 
using the energy loss calculation in standard rock. The total systematic
uncertainty on the expected muon flux, obtained adding in quadrature the errors
from neutrino flux, cross section and muon propagation, is estimated
to be 17 \%. This  uncertainty is mainly a scale error; the error on the shape of the angular distribution is $\sim 5 \%$. 

Fig. \ref{fig6} shows the zenith angle (\(\Theta\)) distribution of the 
measured flux of upthroughgoing
muons. The Monte Carlo expectation for no oscillations is shown as a dashed line. 
Assuming \( \nu _{\mu }\rightarrow \nu _{\tau } \)
oscillations, the best fit parameters are 
\( \Delta m^{2}=2.5\cdot 10^{-3}\, \, \eV ^{2} \) and 
$sin^{2}2\theta = 1$ with a probability of 66\%; the result of the fit is the solid line in Fig. \ref{fig6}.  The probability
for no-oscillations is 0.4 \%. 

\begin{figure}
 \begin{center}
\mbox{\epsfysize=5.5cm
         \epsffile{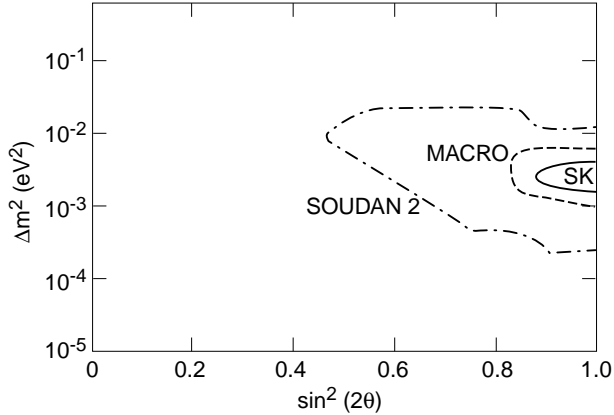}}
\caption{\label{fig7}\small  
 90 \% c.l. allowed regions for $\nu_\mu \rightarrow \nu_\tau$
oscillations from upthroughgoing muons and comparison with the Soudan 2 and 
SK allowed regions (the Soudan 2 region is now smaller).}
\end{center}
\end{figure}

Fig. \ref{fig7}a shows the allowed region for \( \nu _{\mu }\rightarrow \nu _{\tau } \) oscillations in the \( sin^{2}2\theta - \Delta m^{2}\) 
plane, computed according 
to ref. [6] for the upthroughgoing muon events;
our region is compared with those obtained by the SuperKamiokande (SK) [7] and Soudan 2 [8] experiments.

\subsection{Matter effects. $ \nu _\mu \rightarrow \nu _{\tau }$
versus $ \nu _{\mu} \rightarrow \nu _{sterile}$ }

\begin{figure}
  \begin{center}
  \mbox{ \epsfysize=8cm
         \epsffile{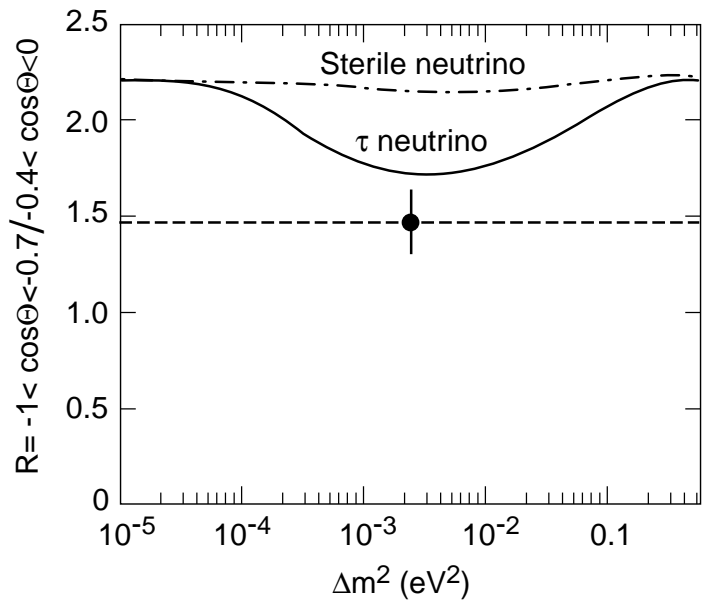} }
\caption{\label{fig8}\small 
Ratio of events with $-1< cos \theta < -0.7$ to events with
$-0.4<cos\theta<0$ vs $\Delta m^2_{23}$ for maximal mixing. The
black point is the measured value, the solid line is the
prediction for $ \nu_\mu \rightarrow \nu_\tau $ oscillations, the dash-dotted
line is the prediction for  $ \nu_\mu \rightarrow \nu_{sterile} $ oscillations.
}
\end{center}
\end{figure}

Matter effects due to the difference between the weak interaction
effective potential for muon neutrinos with respect to sterile neutrinos would produce a different total number
and a different zenith distribution of upthroughgoing muons [9]. 
In Fig. \ref{fig8} the measured ratio between the events with 
$ -1 < cos \Theta < -0.7$ and the
events with $-0.4 < cos \Theta < 0$ is shown as a black point. In this ratio most of the theoretical
uncertainties on neutrino flux and cross section cancel. The remaining
theoretical error combined with the experimental error is estimated to be 
$7\%$. We measured 305 events with $ -1 < cos \Theta < -0.7$ and 206 with $-0.4 < cos \Theta < 0 $; the ratio is R = $1.48 \pm 
0.13_{stat} \pm 0.10_{sys}$. 
For \( \Delta m^{2}=2.5\cdot 10^{-3}\, \, \eV ^{2} \) and maximal
mixing, the expected value of the ratio for 
\( \nu _{\mu }\rightarrow \nu _{\tau } \)
is $R_{\tau} = 1.72$ while for \( \nu _{\mu }\rightarrow \nu _{s} \) is $R_{sterile} = 2.16$.
The maximum probabilities $P_{best}$  to find a value of  $R_{\tau}$ and of $R_{sterile} $ smaller than the expected ones  are 9.4 \% and 0.06 \% 
respectively. The ratio of the maximum probabilities is  
$P_{best_\tau} / P_{best_{sterile} } = 157$, so that \( \nu _{\mu }\rightarrow \nu _{s} \)
oscillations are disfavoured at 99\% c.l. compared
to the \( \nu _{\mu }\rightarrow \nu _{\tau } \) channel [3, 9].

\subsection{$ \ \nu _{\mu }$ energy estimates by MCS of upthroughgoing muons}

\begin{figure}
\vspace{-2.cm}
  \begin{center}
  \mbox{ \epsfysize=8.5cm
         \epsffile{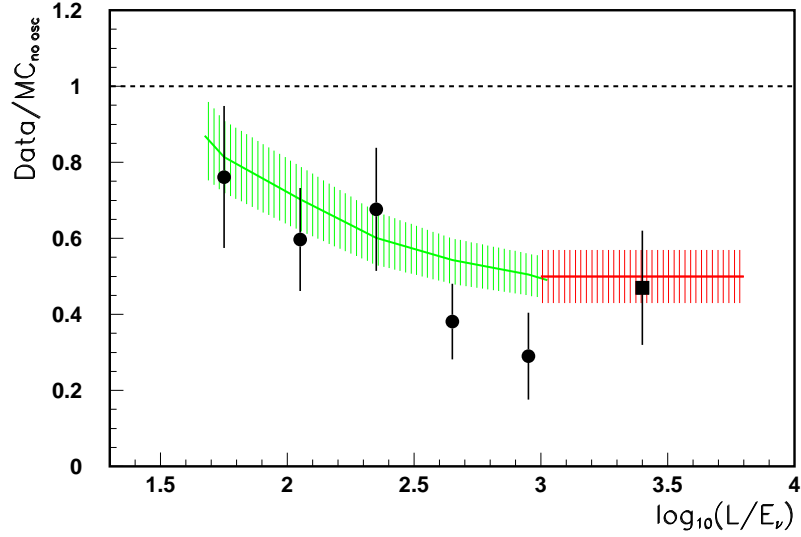} }
\caption{\label{fig9}\small
Data/MC vs $L/E_{\nu}$ for upthrougoing
muons (black circles) and for semicontained up-$\mu$ (black square). The muon energy was estimated by MCS and $E_{\nu}$ by MC
methods. The shaded region represents the uncertainty in the MC
prediction assuming $\sin^2 2 \theta=1$ and $\Delta m^2=0.0025$ eV$^2$. 
The horizontal  dashed line at Data/MC=1 is the expectation for no 
oscillations. 
}
\end{center}
\end{figure}

The oscillation probability is a function of the ratio $L/E_\nu$, Eq. 3. 
Even if a precise measurement of $E_\nu$ is not possible, MC simulations have shown a correlation between $E_\nu$ and the residual muon energy $E_\mu$. 
$E_\nu$ was thus estimated for each event by measuring the muon energy 
$E_\mu$, via Multiple Coulomb Scattering (MCS) of the iduced muon in the absorbers.
The r.m.s. of the lateral displacement for a muon crossing the 
apparatus on the vertical is $\sigma_{MCS}\simeq 10 \cm/ E_\mu (\GeV)$. 
The muon energy $E_\mu$ can be measured up to a saturation
point, occurring when  $\sigma_{MCS}$ is comparable with the detector space resolution.
A first analysis was made studying the deflection of upthroughgoing
muons with the streamer tubes in \lq\lq digital mode''. This method could reach a spatial resolution of $\sim 1 \cm$ which implies a maximum measurable energy  of 10 GeV [10].
As the interesting energy region for atmospheric neutrino oscillations
spans from \( \sim 1\, \GeV  \) to tens of GeV, it is important
to improve the spatial resolution of the detector to push the saturation
point as high as possible. For this purpose, a second analysis was
performed with the streamer tubes used in {}``drift mode{}''. To check the electronics and the feasibility
of the analysis, two tests were performed at the CERN
PS-T9 and SPS-X7 beams. The space resolution achieved is \( \simeq 3{\mm } \), a factor 3.5
better than in the first analysis. For each muon, seven MCS sensitive
variables were given in input to a Neural Network (NN) previously
trained to estimate the muon energy with MC events of known energy crossing the detector
at different zenith angles. The method allowed to separate the upthroughgoing
muons in 4 subsamples corresponding to average neutrino energies  of 12, 20, 
50 and
102 GeV, respectively. The comparison of the 4 zenith angle distributions
with the predictions of the no oscillations MC shows a  disagreement
at low energies (where there is a deficit of vertical events), while
the agreement is restored at the higher neutrino energies.
The distribution of the ratios $R = (Data/MC_{no osc})$ obtained by this
analysis is plotted in Fig. \ref{fig9} as a function of
$log_{10}(L/E_\nu)$ [3] [11]. The black points with error
bars are the data; the vertical extent of the shaded areas represents
the uncertainties on the MC predictions for \( \nu _{\mu }\rightarrow \nu _{\tau } \)
oscillations with maximal mixing and \( \Delta m^{2}=2.5\cdot 10^{-3}\, \,
\eV ^{2} \). The horizontal dashed line is the expectation without oscillations. 
The black square point was obtained from the low energy IU sample.

\subsection{Low energy data.}

\begin{figure}
 \begin{center}
  \mbox{ \epsfysize=6.8cm
         \epsffile{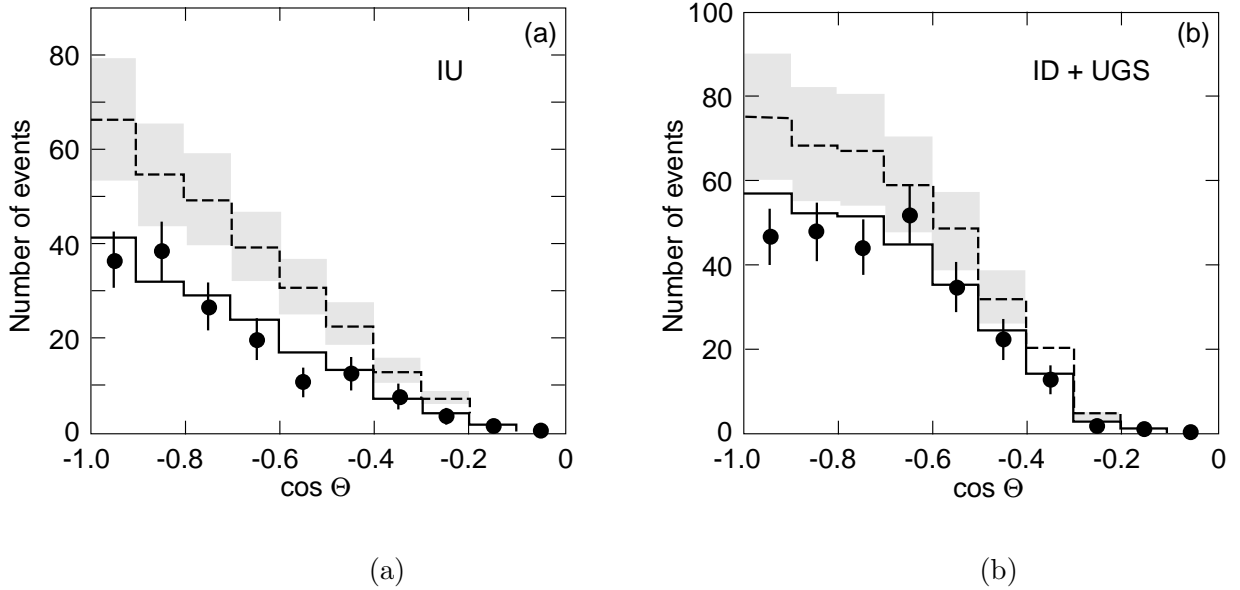}
           }
 \end{center}
{\centering 
{\small \hskip 5.0 truecm (a) \hskip 7.5 truecm (b)}}
\caption{\label{fig10}\small Zenith distributions for (a) the upsemicontained (IU) and 
(b) the upstopping plus the downsemicontained (UGD+ID) events. The black
points are the data, the dashed lines at the center of the shaded regions
correspond to MC predictions 
assuming no oscillations. The full line is the expectation
for  $\nu_\mu \rightarrow \nu_\tau$ oscillations.}
\end{figure}

The \textit{Internal Upgoing (IU) muons} come from \( \nu _{\mu } \)
interactions in the lower apparatus; for these events two scintillation counters
are intercepted; the T.o.F. is applied to identify upward going
muons. The average parent neutrino energy for these
events is 4.2 GeV. For neutrino
oscillations one expects a reduction by about
a factor of two in the flux, without any distortion
in the shape of the angular distribution. This is what is observed in 
Fig. \ref{fig10}a [12].

The \textit{upstopping muons} (UGS) are due to external \( \nu _{\mu } \)
interactions yielding upgoing muons stopping in the detector. The \textit{semicontained downgoing muons} (ID) are due to \( \nu _{\mu } \)-induced
downgoing muon tracks with vertex in the lower MACRO (Fig. 1).
The two types of events are identified by topological criteria; the lack
of time information prevents to distinguish the two sub-samples. An
almost equal number of UGS and ID events is expected.  In case of oscillations,
the flux of the UGS muons should be reduced by 50\%;
 no reduction is instead expected for the semicontained downgoing events
(coming from neutrinos with path lengths of \( \sim 20\km  \)); therefore one 
expects a global reduction of 25\%.

The number of events and the angular distributions are compared with
MC predictions in Table \ref{tab:macro} and Figs. 6a,b. The data show a uniform deficit for the whole angular distribution with respect to 
predictions, \( \sim 50\% \)
for IU, 75\% for ID + UGS; there is good agreement with the predictions
based on neutrino oscillations with the parameters obtained from the
upthroughgoing muons. The average value of the double ratio \( R=(Data/MC)_{IU}/(Data/MC)_{ID+UGS} \)
over the measured zenith angle range is \( R\simeq 0.77 \pm 0.07 \); the
error includes statistical and theoretical uncertainties; \( R=1 \) is expected in case of no oscillations.
\begin{table}
{\centering \begin{tabular}{cccc}
\hline 
                & Events    & MC-No oscillations  & $R = (Data/MC_{no osc}) $\\
                &           &                     &                        \\ \hline
Up throughgoing & $809$ & $1122 \pm 191$ & $(0.721 \pm 0.026_{stat}\pm 0.043_{sys}\pm 0.123_{th} )  $
\\ \hline
Internal Up     & $154$ & $285 \pm 28_{sys}\pm 71_{th}$ & $(0.54 \pm
0.04_{stat} \pm 0.05_{sys} 
\pm 0.13_{th} )$\\ \hline
Up Stop +  In Down &  262 & $ 375 \pm 37_{sys} \pm 94_{th} $ & $( 0.70 \pm
0.04_{stat}\pm 0.07_{sys} \pm 0.17_{th})$\\ \hline
\end{tabular}\par}

\caption{{\small Summary of MACRO $\nu _{\mu}\rightarrow \mu$ events in
    $-1<cos \theta< 0$ (after background subtraction) for oscillation studies.
For each topology (see Fig. \ref{fig2}) the number of measured events, the MC
prediction for no-oscillations and the ratio (Data/MC-no osc) are given [2, 3, 12].
}}

\label{tab:macro}
\end{table}

\begin{figure}
\begin{center}
\vspace{-1cm }
\hspace{-3cm }
  \mbox{ 
\epsfysize=6.6cm
         \epsffile{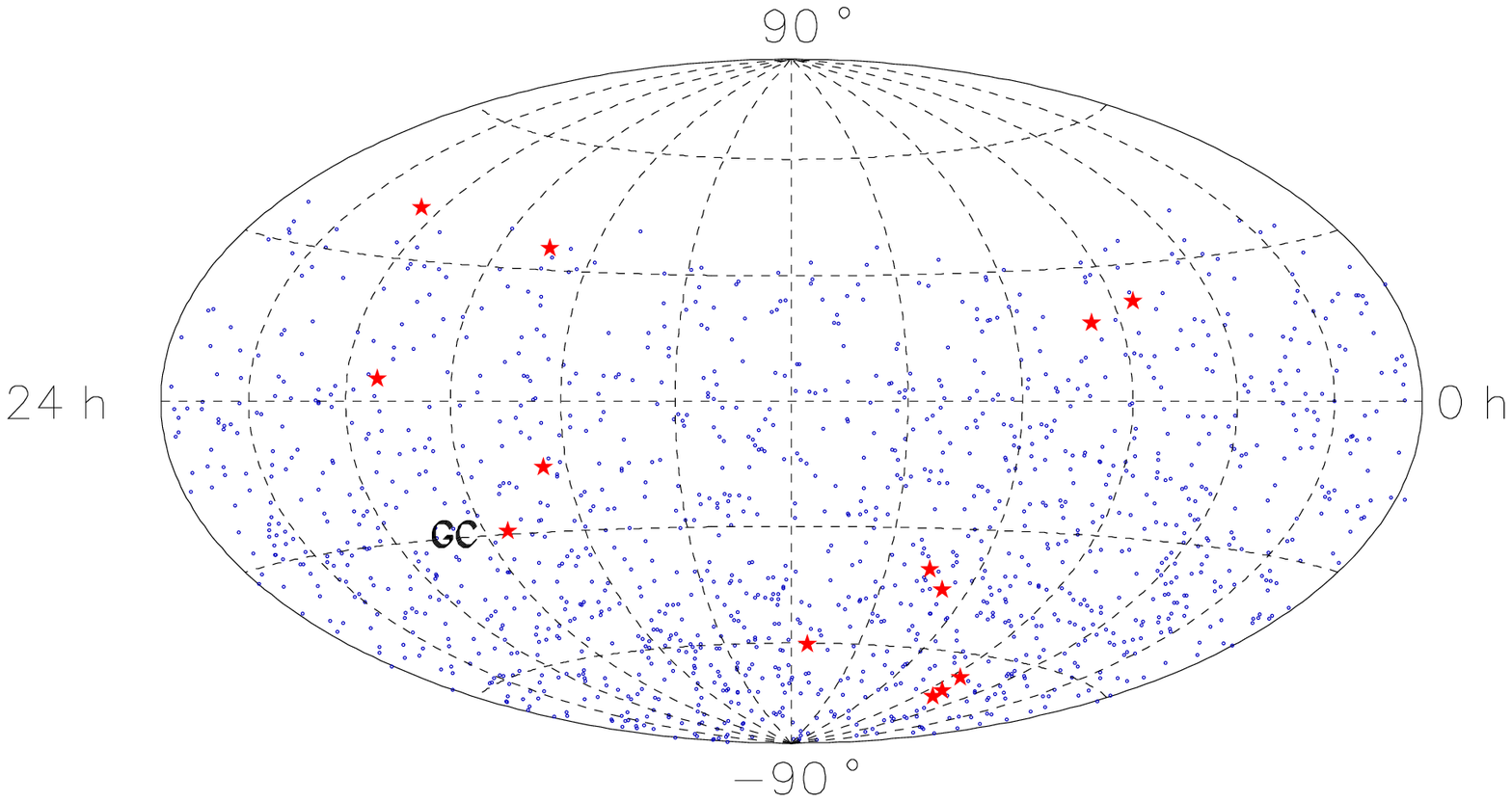}
\hspace{-2.1cm }
\epsfysize=8cm
         \epsffile{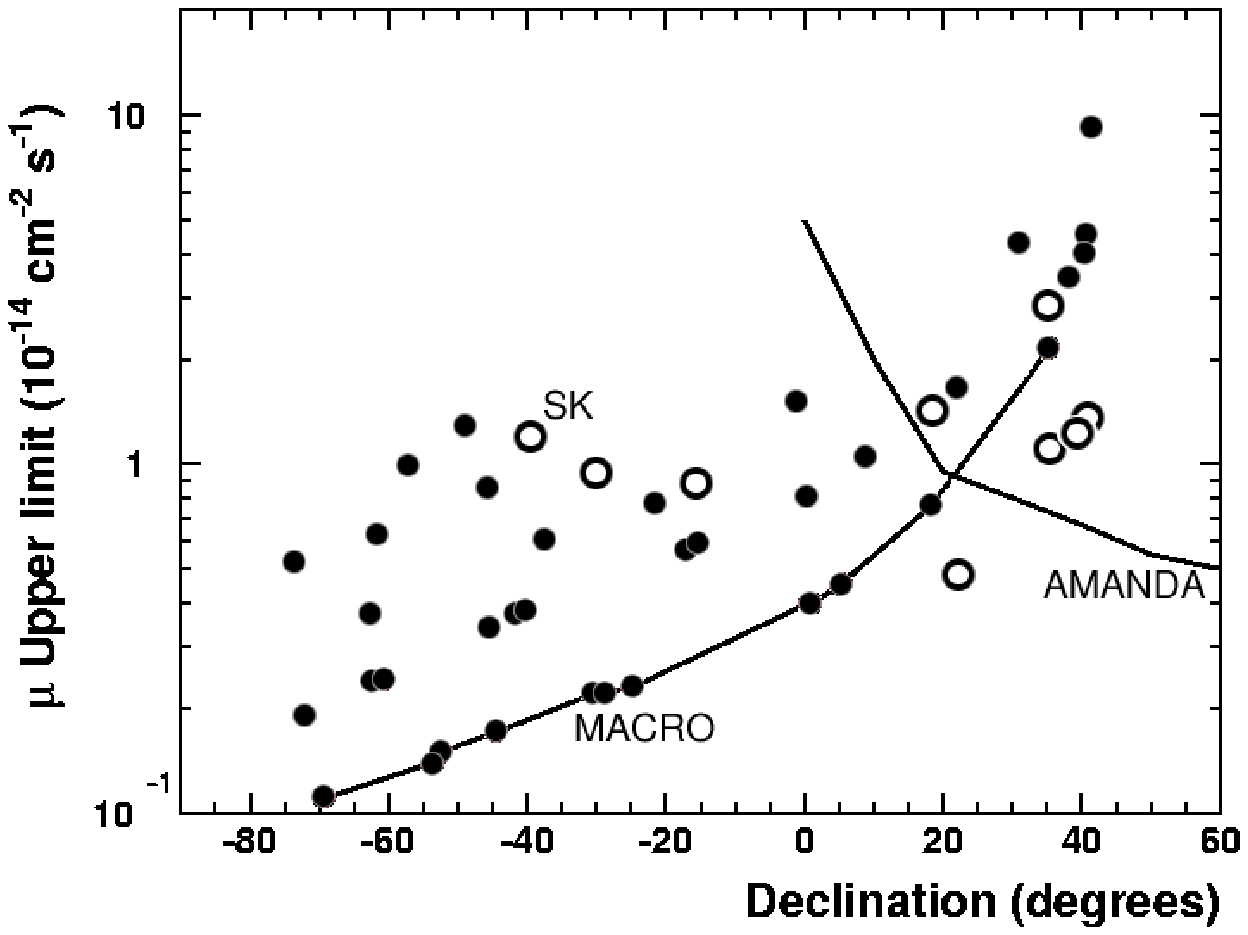} }
{\small \hskip 8.0 truecm (a) \hskip 7.5 truecm (b)}
\caption{\label{fig11}\small 
High energy \(\nu_{\mu}\) astronomy. (a) Upgoing muon distribution in 
equatorial coordinates (1356 events). 
(b) The black points are the MACRO 90 \% c.l. upwardgoing 
muon flux limits as a function 
of the declination for 42 selected sources. The solid line refers to the 
limits obtained for those cases for which the atmospheric neutrino
background was zero. The limits obtained by the SK (open circles) and AMANDA (thin
line) experiments are quoted.
}
\end{center}
\end{figure}

\section{Search for Astrophysical Sources of High Energy Muon Neutrinos}

High energy \( \nu _{\mu } \)'s are expected to come from several
galactic and extragalactic sources. Neutrino production requires
astrophysical accelerators of charged particles and some kind of astrophysical beam
dumps. 

The excellent angular resolution of our detector allowed a
sensitive search for upgoing muons produced by neutrinos coming from
celestial point sources, with a negligible atmospheric neutrino background.
An excess of events was searched for around the positions of known
sources in \( 3^{\circ } \) (half width) angular bins. This value
was chosen so as to take into account the angular smearing due to multiple 
muon scattering in the rock below the detector and by the
energy-integrated angular distribution of the scattered muon, with
respect to the neutrino direction. 

The pointing accuracy of the detector was checked with the observation of the shadow of primary cosmic rays (and thus on downgoing muons) by the Moon and the Sun [3] [13].
It was also checked with the study of the correlations between muons detected 
by MACRO and extensive air showers detected by the EASTOP experiment located above the Gran Sasso massif [14].

In a total livetime of 6.16
y we obtained a total of 1356 events,
see Fig. 7a [3, 13]. The 90\% c.l. upper limits on the muon fluxes from
specific celestial sources lay in the range \( 10^{-15}-10^{-14}{\cm }^{-2}{\s }^{-1} \), see Fig. 7b.
The solid line is our sensitivity vs declination. Notice that we
have two cases, GX339-4 ($\alpha = 255.71^o $, $\delta= -48.79^o $)  and Cir
X-1  ($\alpha = 230.17^o $, $\delta= -57.17^o $), with 7 events: in Fig. 7b 
they are
considered as background, therefore the upper flux limits
are higher; but they could also be indications of signals [3, 15].

We searched for time coincidences of our upgoing muons with sources of 
 \( \gamma  \)-ray
bursts as given in the BATSE 3B and 4B catalogues, for the period
from April 91 to December 2000 [15]. No statistically significant
time correlation was found.

We have also searched for a diffuse astrophysical neutrino flux, for
which we establish a flux upper limit at the level of \( 1.5\cdot
10^{-14}{\cm }^{-2}{\s }^{-1} \) [16].

\section{Indirect Searches for WIMPs}

\begin{figure}
\vspace{-1.cm}
 \begin{center}
  \mbox{ \epsfysize=6cm
         \epsffile{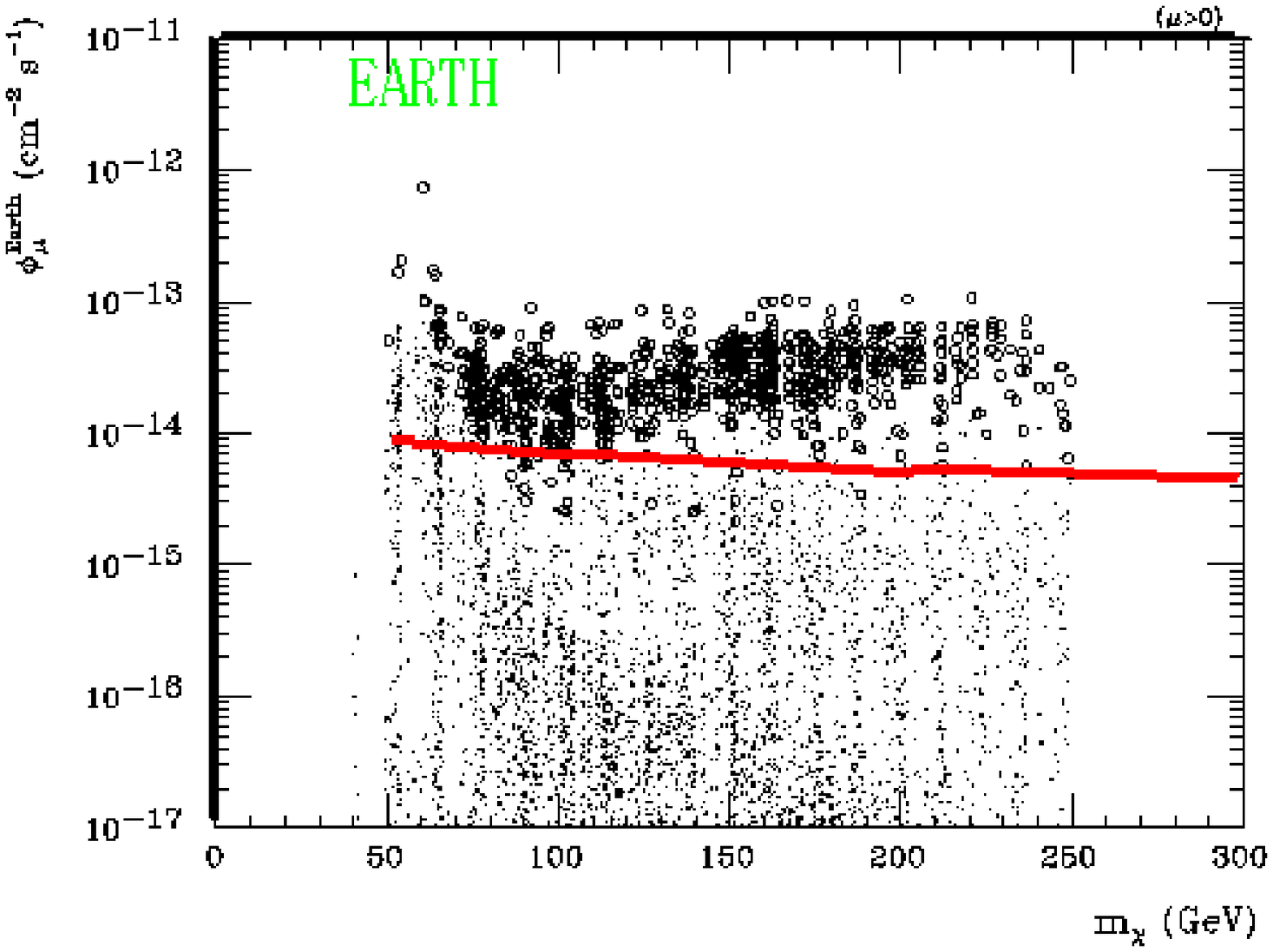}
\epsfysize=6cm
         \epsffile{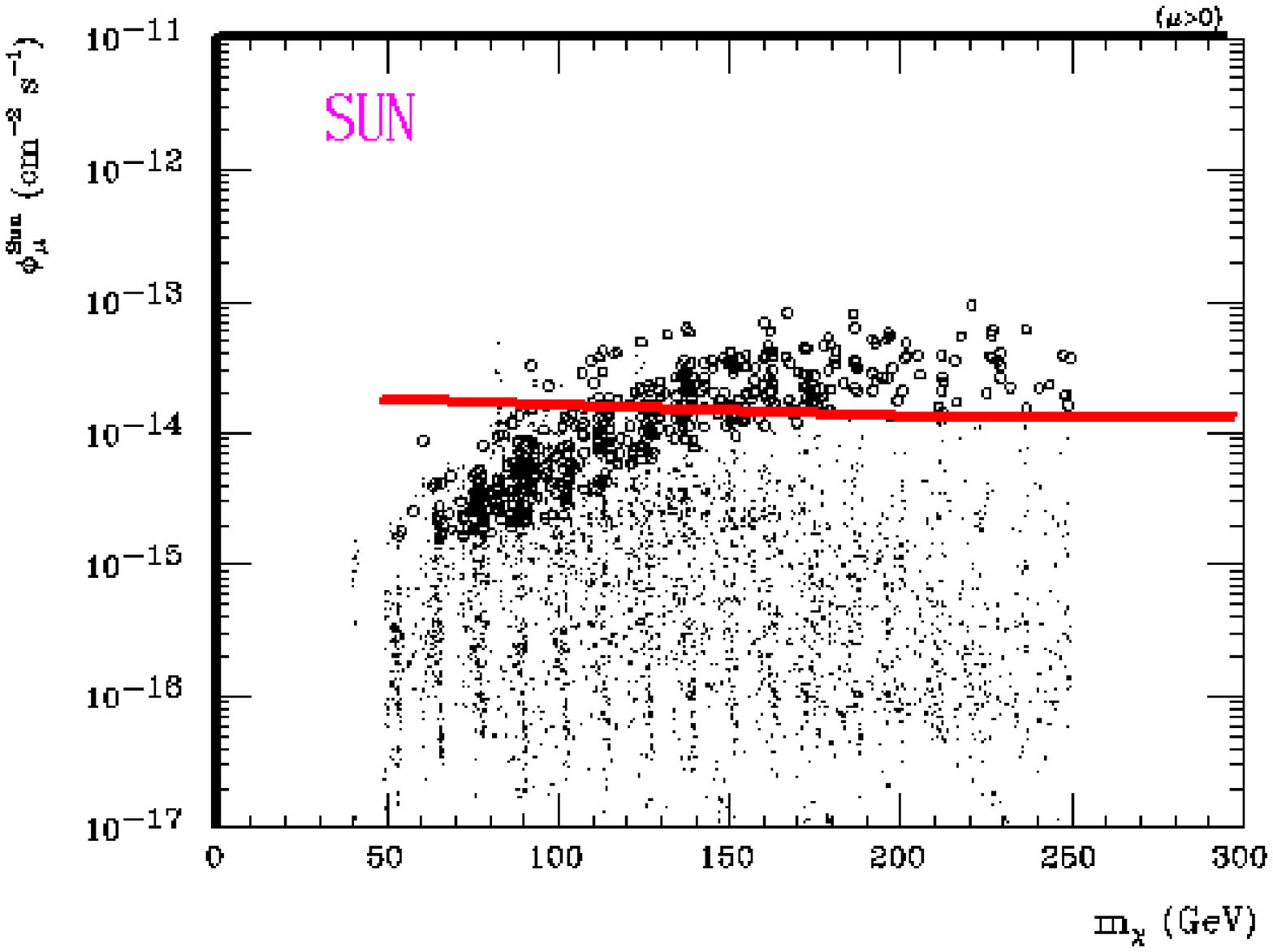}   }
 \end{center}
{\centering 
{\small \hskip 5.0 truecm (a) \hskip 7.5 truecm (b)}}
\caption{\label{fig12}\small  
(a) The solid line is the MACRO upwardgoing muon flux upper limit (90\% c.l.) 
from the Earth plotted
vs neutralino mass $m_\chi$. (b) The same as
in (a) but for upwardgoing muons from the Sun.
Each dot is obtained varying model parameters. The open circles indicate
models {\it excluded  } by direct measurements assuming a local dark matter density of $ 0.5 \GeV cm^{-3}$.}
\end{figure}

Weakly Interacting Massive Particles (WIMPs) could be part of the
galactic dark matter; they could be intercepted by celestial bodies,
slowed down and trapped in their centers, where WIMPs and anti-WIMPs could
annihilate and yield \(\nu_{\mu} \rightarrow \) upthroughgoing muons. The annihilations in these
celestial bodies would yield  neutrinos of \GeV{}-\TeV{} energy,
in small angular windows from their centers. 
For the Earth we have chosen a \( 15^{o} \) cone around the vertical:
we find 863 events. The MC expectation for atmospheric \( \nu _{\mu } \)
without oscillations gives a larger number of events. We set a conservative
flux upper limit assuming that the measured number of events equals the
expected ones. We obtain the \( 90\, \% \) c.l. limits for the upgoing muon flux shown in Fig.  \ref{fig12}a (it varies from 0.8
to 0.5 \( 10^{-14}{\cm }^{-2}{\s }^{-1} \)) [17, 3]. If the WIMPs are 
identified with the smallest mass neutralino, our limits may be used to
constrain the neutralino mass, following the model of ref. [18], 
Fig. \ref{fig12}a.

A similar procedure was used to search for muon neutrinos from the
Sun, using 10 search cones from \( 3^{o} \) to \( 30^{o} \). In
the absence of statistically significant excesses the muon flux upper limits
are at the level of \( 1.5-2\cdot 10^{-14}{\cm }^{-2}{\s }^{-1} \).
The limits are shown in Fig. 8b as a function of the WIMP (neutralino) mass.

\section{Neutrinos from Stellar Gravitational Collapses}

A stellar gravitational collapse (GC) of the core of a massive star
is expected to produce a large burst of all types of neutrinos and
antineutrinos with energies of \( 7-30 \)~\MeV{} and with a duration
of \(<10{\s } \). The \anue{}'s can be detected via the process
\( \bar{\nu }_{e}+p\rightarrow n+e^{+} \) in the liquid scintillator.
About \( 100\div 150 \) \anue{} events should be detected in our
580 t scintillator for a stellar collapse at the center of our Galaxy.

We used two electronic systems to search for bursts of \anue{}'s from stellar
gravitational collapses. The first system was based on the dedicated
PHRASE trigger, the second one was based on the ERP general trigger. Both
had energy thresholds of \( \sim 7{\MeV } \) and recorded
pulse shape, charge and timing informations. Immediately after a 
trigger, the PHRASE system lowered its threshold to 1 MeV, for
a duration of \( 800{\mu \s } \), in order to detect (with a \( \simeq 25\, \% \)
efficiency) the \( 2.2{\MeV } \) \( \gamma  \) released in the reaction
\( n\, +\, p\rightarrow d\, +\, \gamma _{2.2\MeV } \) induced by
the neutron produced in the primary process.

A redundant supernovae alarm system was in operation, alerting immediately
the physicists on shift. We defined a general procedure to alert the
physics and astrophysics communities in case of an interesting alarm. Finally, a procedure to link the various supernovae
observatories around the world was set up [3, 19].

The effective MACRO active mass was \( \sim 580\, \,  \)t; the live-time fraction was \( \simeq 97.5\, \% \); we kept always ruming at least 1-2 supermodules. No stellar gravitational
collapses were observed in our Galaxy from the beginning of \( 1989 \)
to the end of 2000 [3][19].

\section{Conclusions}

The MACRO detector took data from 1989 to the end of year 2000 and it obtained important results in all the items listed in the proposal, in particular:
\begin{itemize}

\item Atmospheric neutrino oscillations. Analyses of different event 
topologies at different
energies, the exploitation of the muon Coulomb multiple scattering in the detector
give strong support to the hypothesis of $\nu_\mu \rightarrow \nu_\tau$ oscillations with  \( \Delta m^{2}=0.0025 eV^{2}\) and maximal mixing. The statistical and systematic significance of our global data is well above at the 5 
standard deviation level. We are at present performing some global analyses 
using also the new data on atmospheric muon flux measurements.
\item High energy muon neutrino astronomy. MACRO has been highly competitive
with other underground experiments thanks to its large acceptance and good angular accuracy.
It has been limited by its livetime and the size of the detector.
\item Search for bursts of $\bar{\nu}_e$ from stellar gravitational
collapses. MACRO was sensitive to supernovae events in our Galaxy,
it started  the SN WATCH  system, and for a certain time it was the only detector in operation.
\item Sensitive indirect searches have been carried out for possible Dark Matter 
candidates like the neutralinos, looking
for upgoing muons  from the center of the Earth and from the Sun.

\end{itemize}

It may be pointed out that MACRO also obtained the best flux upper limits for 
GUT Magnetic Monopoles over the widest $\beta$ range [20].

\subsection*{Acknowledgments}
I thank the members of the MACRO Collaboration for their cooperation; in particular I acknowledge the discussions and the help from the members of the Bologna group, in particular Drs. Y. Becherini and M. Giorgini. I thank Ms Anastasia Casoni for typing and correcting the manuscript.


\end{document}